# Sampling-based probabilistic inference emerges from learning in neural circuits with a cost on reliability

Laurence Aitchison[1]*, Guillaume Hennequin[1], Máté Lengyel[1,2]

[1] Computational & Biological Learning Lab, Department of Engineering, University of Cambridge, Cambridge, United Kingdom

[2] Department of Cognitive Science, Central European University, Budapest, Hungary

* laurence.aitchison@gmail.com

## Abstract

Neural responses in the cortex change over time both systematically, due to ongoing plasticity and learning, and seemingly randomly, due to various sources of noise and variability. Most previous work considered each of these processes, learning and variability, in isolation – here we study neural networks exhibiting both and show that their interaction leads to the emergence of powerful computational properties. We trained neural networks on classical unsupervised learning tasks, in which the objective was to represent their inputs in an efficient, easily decodable form, with an additional cost for neural reliability which we derived from basic biophysical considerations. This cost on reliability introduced a tradeoff between energetically cheap but inaccurate representations and energetically costly but accurate ones. Despite the learning tasks being non-probabilistic, the networks solved this tradeoff by developing a probabilistic representation: neural variability represented samples from statistically appropriate posterior distributions that would result from performing probabilistic inference over their inputs. We provide an analytical understanding of this result by revealing a connection between the cost of reliability, and the objective for a state-of-the-art Bayesian inference strategy: variational autoencoders. We show that the same cost leads to the emergence of increasingly accurate probabilistic representations as networks become more complex, from single-layer feed-forward, through multi-layer feed-forward, to recurrent architectures. Our results provide insights into why neural responses in sensory areas show signatures of sampling-based probabilistic representations, and may inform future deep learning algorithms and their implementation in stochastic low-precision computing systems.

## Introduction

Neural circuits change the way they respond to stimuli on multiple time scales. On slow time scales, synapses, dendrites, and cells undergo plasticity, changing their molecular, morphological and electrophysiological properties [1]. These changes offer powerful substrates for optimising task performance [2, 3, 4, 5, 6], and have indeed been shown to underlie learning in a diverse set of behavioural paradigms [7, 8]. On fast time scales, neural responses exhibit substantial variability both within and across trials, with different numbers of vesicles being released upon the arrival of each presynaptic spike at the level of individual synapses [9], and different spiking responses evoked by the same stimulus at the level of neural populations [10, 11]. This response variability often predicts trial-by-trial fluctuations in behaviour, such that particular patterns of neural responses are associated with worse performance than others [12, 13, 14]. Thus, the question arises: if plasticity offers such powerful ways to optimise neural responses, and variability is detrimental to performance, why do circuits not learn to entirely suppress variability? More importantly, what kind of trade-offs exist between plasticity and stochasticity and how are these trade-offs reflected in neural responses?

To study the interaction between variability and plasticity, we trained neural network models to improve their performance on standard unsupervised learning problems. Critically, we took into account that making responses reliable is costly and formalised this cost based on basic biophysical considerations. Thus, by treating variability as being subject to plasticity [9, 15] alongside all other parameters affecting performance, we were able to study the tradeoff between maximising task performance and minimising biophysical costs.

We studied this tradeoff in a variety of network architectures of increasing complexity. We first investigated the effects of plastic variability in a classical sparse-coding model that was trained to develop sparse representations that afforded an accurate reconstruction of its inputs (natural image patches) [6]. We found that optimised networks retained finite amounts of variability that was shaped such that responses preferentially "sampled" regions of the state space associated with lower sparse coding cost. In order to understand the functional significance of these sculpted response distributions, we used a probabilistic interpretation of the sparse coding model as using an internal model of natural images to infer the hidden causes underlying its input [16]. This probabilistic interpretation revealed a stronger result about the response distributions of the model: the optimised networks were performing approximate Bayesian inference by sampling the posterior distributions of the internal model [17, 18, 19, 20, 21]. We demonstrate the generality of these results by showing analytically that our biologically motivated objective forms an upper bound on a statistically principled objective, the variational free-energy, used to train variational autoencoders in modern machine learning applications [22, 23, 24, 25]. As a consequence, optimizing our objective always increases the similarity between response distributions of the network and the probabilistic posteriors of the corresponding internal model.

We then extended these results to a deep neural network that was able to learn a more complex internal model of its inputs (handwritten digits). We also studied recurrent networks and showed that they matched their response distributions to complex posterior distributions with correlations – a task on which feed-forward architectures inherently perform poorly. While recurrent dynamics came with the drawback of introducing temporal correlations in the population response (i.e. between consecutive samples), making inference slow, this could be mitigated by explicitly encouraging fast convergence without affecting autoencoder performance and the accuracy of inference. These results suggest that sampling-based probabilistic inference arises naturally, or even unavoidably, and thus may be a general computational strategy adopted by sensory circuits operating under biophysical constraints.

# 1 Results

## 1.1 Sparse coding with an energetic cost on reliability

To study the interaction between variability and learning, we first considered a classical model of primary visual cortex, the sparse coding model [6]. This model consists of a single layer of neurons, with responses $\mathbf{z}$, which receive sensory input data in the form of an image patch, $\mathbf{x}$ (Fig. 1A), through a set of feed-forward synaptic weights, $\mathbf{W}$ (Fig. 1A):

$$z_i(\mathbf{x}) = \sum_j W_{ij}\, x_j + \sigma_i\, \eta_i \tag{1}$$

where $\eta_i$ is standard Gaussian noise, scaled by a neuron-specific parameter $\sigma_i$. We trained the network to form an accurate representation of its input $\mathbf{x}$, such that $\mathbf{x}$ could be reconstructed from the responses $\mathbf{z}$ with high fidelity as a linear readout $\hat{\mathbf{x}}$ with reconstruction coefficients $\mathbf{A}$ (that is, $\mathbf{x} \approx \hat{\mathbf{x}} = \mathbf{A}\mathbf{z}$), while also maintaining sparse neural responses. Formally, the training objective (the "sparse coding cost") consisted of the squared error between the original image and the reconstruction, and a sparseness cost on neural responses:

$$\text{sparse coding cost} = \underbrace{\text{reconstruction error}}_{||\mathbf{x}-\hat{\mathbf{x}}||_2} + \underbrace{\text{sparseness cost}}_{||\mathbf{z}||_1} \tag{2}$$

and we tuned all parameters $\mathbf{W}$, $\mathbf{A}$, and $\boldsymbol{\sigma}$ jointly. While optimising $\mathbf{W}$ and $\mathbf{A}$ is possible using known synaptic plasticity rules [26, 27, 28, 29, 30, 31, 32, 33], we focused on results that are generic to all such plasticity rules, by considering the *outcome* of optimisation rather than the detailed dynamics by which $\mathbf{W}$ and $\mathbf{A}$ reach their optimum. The tuning of $\boldsymbol{\sigma}$ is more unusual (though see [9] for possible biological substrates) and we chose this parametrisation for ease of exposition in the feedforward setting. Importantly, as we will show later in a more realistic (recurrent) setting, the *effective* variability of neurons can in fact be optimised by changing the (recurrent) synaptic weights, without introducing explicitly tunable $\boldsymbol{\sigma}$ variables.

Note that the feed-forward architecture of our model was considerably simpler than previous sparse coding models which also included recurrent connections between the neurons. This allowed us to avoid the "weight-transport problem" [35] of classical recurrent sparse coding architectures, in which both recurrent and feedforward weights needed to be constrained based on the reconstruction coefficients rather than learnt [6, 16]. Thus, we could learn the feed-forward synaptic weights, $\mathbf{W}$, and reconstruction coefficients, $\mathbf{A}$, at the same time. In spite of its simplicity, when trained on whitened natural images (Fig. 1C), neurons of the network developed receptive fields that were selective to oriented edge-like features (Fig. 1D) similar to the standard sparse coding model and simple cells of the primary visual cortex [36, 37]. However, given that variability can only impair reconstruction performance (and also increase the sparseness penalty), the network also learnt to have extremely reliable responses (or, equivalently, very low variability, $\sigma_i$; Fig. 1E-F). Real neural circuits display finite variability, so we considered the biophysical factors that prevent neural systems from eliminating variability while optimising their performance, and studied the surprising computational consequences of these factors on the emergent behavior of the system.

A fundamental reason for why neural circuits cannot achieve infinite reliability is that reliability comes at an energetic cost. To understand the exact form of this reliability cost, we consider the underlying biophysics [38]: neurons rely on intrinsically "stochastic elements" (e.g. ion channels, released synaptic vesicles) and so in order to increase reliability, neural systems need to average together a larger number of these elements. Consider a network in which each unit, $z_i$, averages across $N_i$ stochastic elements (with limited correlations). The standard deviation of this average scales as $\sigma_i \propto 1/\sqrt{N_i}$, and we can expect the energetic cost associated with each unit to scale linearly with $N_i$. Thus the total energetic cost in the network, $C$, is going to scale with the total number of stochastic elements, i.e. as the sum of inverse variances (Fig. 1B):

$$\text{reliability cost} \propto \sum_i N_i \propto \sum_i \frac{1}{\sigma_i^2} \tag{3}$$

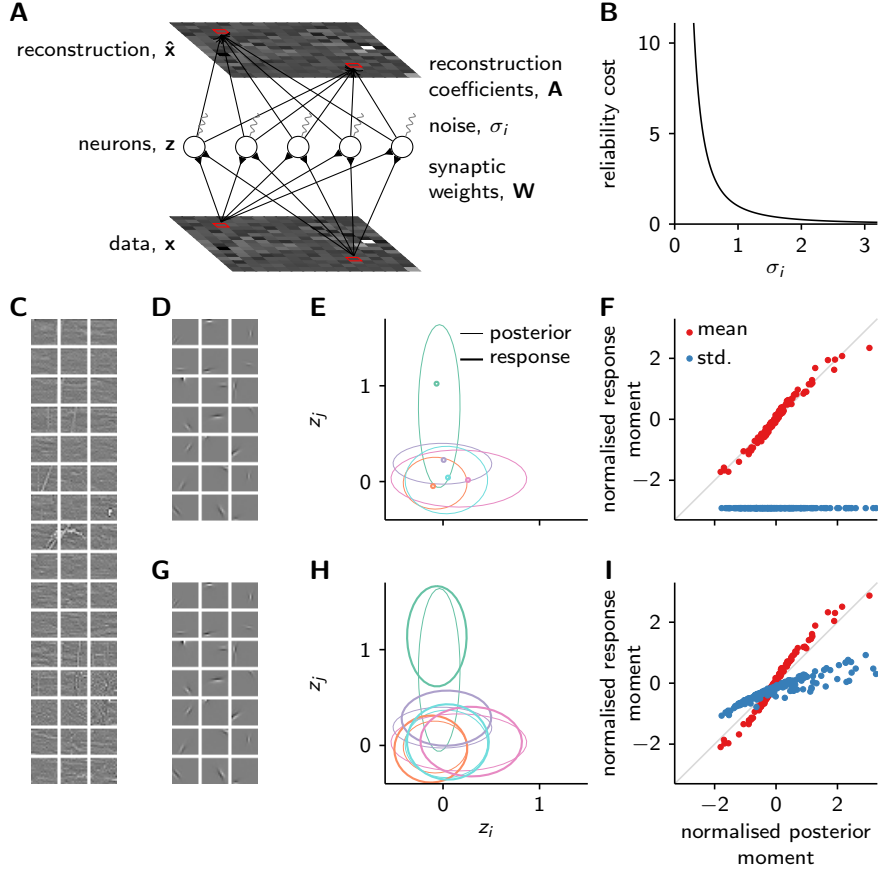

**Figure 1:** Sparse coding model with variability. **A.** Schematic diagram of the network, with input $\mathbf{x}$ being a $16 \times 16 = 256$-pixel grayscale image, $\mathbf{z}$ being the responses of 192 neurons (as in [6]), receiving input through their feed-forward synaptic weights, $\mathbf{W}$, and subject to noise, $\sigma_i$. The reconstructed input, $\hat{\mathbf{x}}$, is computed as a linear combination of neural responses with coefficients $\mathbf{A}$. **B.** Dependence of the reliability cost on the variance of a single unit (Eq. 3). **C.** Whitened natural images [34] on which the network is trained. **D.** Learned feedforward synaptic weights, $\mathbf{W}$, without reliability cost. **E.** Covariance ellipses displaying example neural response distributions (thick lines) and the corresponding true posteriors (thin lines) for a representative input. Different colors correspond to different pairs of cells. Note that the response distributions are very narrow, so they appear as points. **F.** Mean (red) and standard deviation (blue) of the posterior (x-axis) and response distributions (y-axis), normalized by the mean and standard deviation of the relevant posterior moment. Different dots correspond to different cells. **G-I.** As **D-F**, but incorporating the reliability cost. Note that response distributions have non-zero standard deviations which match those of the corresponding posteriors.

This reasoning applies to a range of different scenarios (for more details, see Methods Sec. 3.3). First, when each $z_i$ is encoded by a single cell, the signal-to-noise ratio can be improved by averaging over more spikes, i.e. by increasing the firing rate (while decreasing the outgoing weights to ensure that the expected total downstream activation is left unchanged). Second, when each $z_i$ is encoded by a distinct "sub-population" of cells, the signal-to-noise ratio can be improved by increasing the number of cells in these sub-populations. Third, when variables $z_i$ are linearly embedded in the responses of a larger population of cells, so there is no need for distinct sub-populations, the signal-to-noise ratio can be improved by increasing the overall number of cells in the population. Alternatively, whichever representational scheme is used, recurrent networks can also suppress noise by jointly scaling up inputs and recurrent connection strengths, which could be achieved by scaling up the number of ion channels. Interestingly, while this strategy does not explicitly rely on averaging, it turns out to yield the same scaling between energy and reliability.

We therefore retrained the sparse coding network with a modified cost function that included this reliability-dependent energetic cost:

$$\text{total cost} = \text{sparse coding cost} + \text{reliability cost} \qquad (4)$$

We found that the network developed receptive fields that were very similar to those of the deterministic network (Fig. 1G) but, as expected, neural responses retained a finite variability (Fig. 1H-I).

## 1.2 Differential variability and the trade-off between sparse coding and reliability costs

Not only did the cost on reliability prevent variability from falling to zero, the resulting optimized variability exhibited a particular structure: different cells exhibited different amounts of variability (Fig. 1H-I) despite the explicit reliability cost being the same across cells (Eq. 3). To understand the source of this structure, we compared it with the simpler cases when there was very low variability in the network (essentially equivalent to the original network we trained without a reliability cost), or when variability was larger but the same across cells and was not optimized jointly with other parameters (Fig. 2). If we take the noise to be very low, the neural network finds the single best encoding of the input (Fig. 2A, left), and indeed achieves a very low sparse-coding cost (which includes both the reconstruction error and the sparseness cost; Fig. 2B, red hatched). However, the reliability cost is very high (if the noise were zero, the penalty would be infinite), overriding those gains (Fig. 2B, solid red). One way to reduce the reliability penalty is to incorporate high, fixed noise (Fig. 2A, middle). While this does yield a much smaller reliability cost (Fig. 2B, solid blue), the sparse coding cost has increased considerably, as reconstruction errors are inevitably larger if there is more variability in the neural representation (Fig. 2B, blue hatched). Moreover, the increase in reconstruction error associated with variability differs across directions, being larger in the $z_1$ direction than the $z_2$ direction in Fig. 2A. Thus, when we optimize the variability along each direction, the system finds the optimal way to balance the trade-off between sparse coding and reliability costs, resulting in cell-specific variability. In this example, variability along the $z_1$ direction decreases, to avoid the large reconstruction error penalty, while the variability along the $z_2$ direction can be larger to incur a smaller reliability cost, as deviations in this direction matter less (Fig. 2A right). Resolving these detailed trade-offs means that the resulting network has a substantially lower variability cost than the low-noise regime, whilst keeping the sparse-coding cost (driven by the reconstruction error along some, but not all directions) under control (Fig. 2B, green).

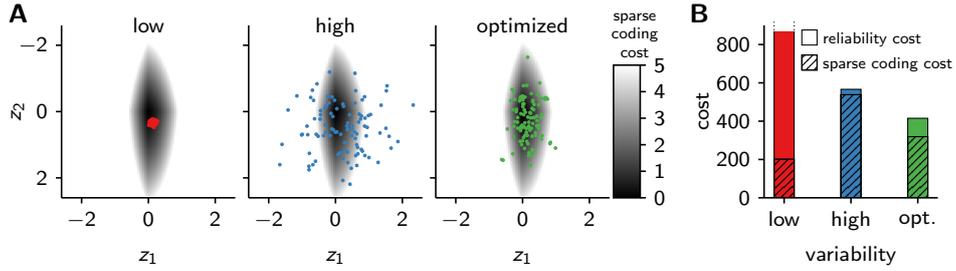

**Figure 2:** Differential variability emerges as a result of the trade-off between sparse coding and reliability costs. **A.** Samples of network activity (coloured dots) for a fixed input from networks under different constraints on variability. A state space of two neurons ($z_1$ and $z_2$) is shown. From left to right: low level of fixed variability (red), high level of fixed variability (blue), level of variability jointly optimized with other parameters (green). Gray-scale maps show the sparse coding cost as a function of network state (keeping all other neuronal activities fixed), which was the same in all networks. **B.** Total costs for each network (bars, colors as in A), combine the sparse coding cost (hatched segments) and the reliability cost (solid segments).

## 1.3 Link between optimized variability and Bayesian inference

The detailed matching of variability to the geometry of the sparse coding cost in the jointly optimized network led to neural activities appearing as if they were sampling a distribution defined by the region of low sparse coding cost (Fig. 2A, green dots vs. gray scale map). To further investigate the potential link between optimized variability and a probabilistic representation, we exploit a well-known probabilistic interpretation of the sparse coding cost [39]. Under this interpretation, the sparse coding cost reflects a probabilistic internal model of the process that generates the input. According to this internal model, inputs arise as a linear combination of some latent causes in the environment, with additive Gaussian noise (Gaussian likelihood, $P(\mathbf{x}|\mathbf{z})$), where these latent causes are sparsely distributed (Laplacian prior, $P(\mathbf{z})$). Noting the one-to-one correspondence between the components of the sparse coding cost and the components of this probabilistic model, reconstruction error $= -\log P(\mathbf{x}|\mathbf{z}) + \text{const}$ and sparseness cost $= -\log P(\mathbf{z}) + \text{const}$, we can rewrite the sparse coding cost in probabilistic terms as

$$\text{sparse coding cost}(\mathbf{z}) = -\log P(\mathbf{x}|\mathbf{z}) - \log P(\mathbf{z}) + \text{const} = -\log P(\mathbf{z}|\mathbf{x}) + \text{const} \quad (5)$$

Thus, the sparse coding cost is equivalent to the (negative logarithm of the) posterior probability that a particular setting of latent causes might account for the input, $P(\mathbf{z}|\mathbf{x})$. Therefore, minimising the sparse coding cost amounts to finding the most probable causes of the input under this probabilistic model.

However, as we saw above, the additional reliability cost in the optimized stochastic network prevents neural activities from deterministically finding the minimum of the sparse coding cost (and thus the maximum of the posterior probability). Instead, neural responses display variability, which is shaped by the sparse-coding cost so as to preferentially sample low-cost regions, or equivalently (given Eq. (5)) regions of high posterior probability. Importantly, preferentially sampling regions of high posterior probability is exactly what we would expect under an (approximate) inference scheme, in which neural activity samples from the posterior distribution over latent causes, conditioned on sensory input. Indeed, a direct comparison of the posterior distributions of the probabilistic model and the response distributions of the optimized stochastic network show a close match (Fig. 1H-I), suggesting that for each input the dynamics of the network produce samples from the corresponding posterior distribution.

To formally understand the link between optimized variability and sampling posterior distributions we noted a close correspondence between our network and a state-of-the-art machine learning method for performing approximate probabilistic inference in neural networks, variational autoencoders [24, 25]. The objective in variational autoencoders, the variational free energy ($\mathcal{F}$), is very similar to the total cost of our network:

$$\text{total cost}(\mathbf{z}) = -\log P(\mathbf{z}|\mathbf{x}) + \text{reliability cost} + \text{const} \tag{6a}$$

but where our biologically-motivated reliability cost is replaced by the negative entropy of the response distribution,

$$\mathcal{F}(\mathbf{z}) = -\log P(\mathbf{z}|\mathbf{x}) - \text{entropy} \quad + \text{const} \tag{6b}$$

Thus, both objectives can be written as a sum of two terms: one that prefers neural responses that represent latent variables with high posterior probability and another that encourages the response distribution to spread out. Although the latter term is not equivalent in the two expressions, we can show (Methods Sec. 3.2) that the biophysically motivated reliability cost of our network approximates and forms an upper bound on the usual variational free-energy. As it is known that minimizing the variational free energy results in posterior samplers [22], this implies that our network will also perform approximate inference by drawing samples from the posterior distribution, explaining the close match between the posterior and response distributions that we found in our model (Fig. 1H-I).

Furthermore, one of the key properties of the variational free-energy is that it upper bounds the negative "model evidence", $-\log P(\mathbf{x})$ [40], and by combining these two bounds, we see that our total cost upper-bounds the variational free energy, which upper-bounds the negative model evidence,

$$-\log P(\mathbf{x}) \leq \mathrm{E}_\mathbf{z}\left[\mathcal{F}(\mathbf{z})\right] \leq \mathrm{E}_\mathbf{z}\left[\text{total cost}(\mathbf{z})\right] \tag{7}$$

Thus, by optimizing our biological objective, we ultimately optimize the model evidence ensuring that the implicit internal probabilistic model which is assumed to generate the input will be well-calibrated.

## 1.4 Multi-layered architecture

Having established the formal relationship between our neural networks, with a biologically motivated reliability cost, and a well-understood machine learning method, variational inference, we were able to extend the approach to a multi-layered architecture. In particular, we implemented a multi-layered model similar to the sparse coding model, but with a hierarchy of 5 feed-forward connected layers, with 64 neurons in the first (lowest) layer, 32, 16, 8, and 4 in the last (highest) layer (Fig. 3A), and with each neuron having 40 non-linear dendrites (Fig. 3B), which receive input from all cells in the preceding layer [41, 42, 43, 44]. We then trained our network on a standard unsupervised learning benchmark, the MNIST handwritten digit data set [45], using a reconstruction error analogous to that used in the sparse coding model (Fig. 3C, Methods Sec. 3.6). To check the quality of the learned model, we generated images from the probabilistic internal model defined by the reconstruction coefficients (Fig. 3C). While the resulting samples did not match state-of-the-art machine learning methods (e.g. [46]), they were nonetheless remarkable given that the objective uses a biologically-motivated reliability cost, rather than the theoretically-motivated entropic reliability cost. Furthermore, the response distributions again matched the true posteriors (Fig. 3E).

Examining the response distributions of our network more closely revealed that the response and true posterior means were tightly correlated as we changed the input image, but the network had more difficulty modulating the standard deviation and correlations of responses across images (Fig. 3F-G). As a reference, note that for the sparse coding model we demonstrated cell- but not input-dependence of the standard deviation of responses (Fig. 1H-I). Modulating the variability of responses in an input-dependent manner is a particularly challenging task for a network [47] – which we could not study in the sparse coding model both because its posteriors showed minimal correlations (and modulation of these correlations, Fig. 1E, H) and because the neural network we used was also too simplistic. Remarkably, some (albeit limited) input-dependent modulation of variability emerged in the multi-layer network, despite the fact that, as in the sparse coding model (Eq. 1), neural responses conditioned on input from the preceding layer remained independent (i.e. have zero correlations) with fixed standard deviations. However, in the multi-layer network, independent variability in lower layers created input-dependent, correlated variability in higher layers as it propagated through the synaptic connections and neural firing-rate nonlinearities of the network. While these effects did build up as activity propagated to higher layers (Fig. 3H), they were inherently limited as this architecture could not allow neural response correlations and standard deviations to depend directly on neural activities in the preceding layer.

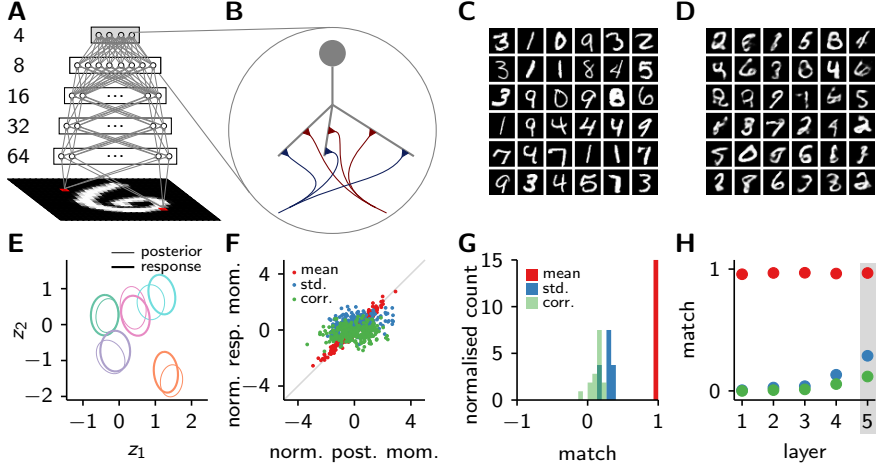

**Figure 3:** A hierarchical neural network with a reliability cost, trained to reconstruct MNIST handwritten digits. **A.** Schematic diagram showing the architecture of the model, the data is a $28 \times 28 = 784$-pixel binarised image of a handwritten digit, followed by five layers of 64, 32, 16, 8, and 4 units, respectively. The reconstruction coefficients are omitted for clarity. **B.** Schematic of a single neuron, showing the inputs from the previous layer, which impinge on 40 separate nonlinear dendrites, and finally are combined to give the output. **C.** Data images from the MNIST database of handwritten digits used to train the network. **D.** Images generated by our model after training. **E.** Covariance ellipses corresponding to the top-layer response distributions (thick lines) and posterior distributions (thin lines) for a representative pair of cells. Different colors correspond to different images. Note that despite each layer being independent conditioned on the previous layer, we do see some correlations in the response distributions (tilted ellipses), due to integrating across different possible settings in the lower layers. **F.** Mean (red), standard deviation (blue), and correlation (green) of the top-layer posterior (x-axis) and response distributions (y-axis), normalized by the mean and standard deviation of the relevant posterior moment. Different dots correspond to different images. **G.** Histogram of correlation coefficients "match" from **F** across cells and cell-pairs. **H.** Correlation coefficient "match" between posterior and response distribution moments across images, averaged across cells or cell-pairs (colors as in E). Note that the matched input-dependent modulation of the standard deviation and correlation slowly increases as noise is passed through the network.

## 1.5 Linear recurrent network

While cortical networks do perform feedforward processing (e.g. when rapidly transforming visual stimuli into actions in the face of synaptic and action potential delays [48]), they also have a high degree of recurrent connectivity. In order to study the effects of recurrent processing in the context of unsupervised learning, we revisited the sparse coding model by supplementing the feed-forward mapping from the input to the neural responses with recurrent connections among the neurons (Fig. 4A). To a first approximation, the consequences of recurrent processing can be understood by considering the dynamics of the following standard, highly simplified (linear) network model [49, 50, 51, 52]:

$$\dot{z}_i = \frac{1}{\tau} \left( -z_i + \sum_j W_{ij}^{\text{rec}} z_j + \sum_k W_{ik} x_k \right) + \frac{\sigma}{\sqrt{\tau}} \eta_i \qquad (8)$$

where $\tau = 100$ ms is the neural time constant, $W_{ij}^{\text{rec}}$ is the weight of the recurrent synapse going from neuron $j$ to neuron $i$ ($W_{ii}^{\text{rec}} = 0$, i.e. self-connections are not allowed), and $\mathbf{W}$ and $\boldsymbol{\eta}$ respectively represent the feed-forward weights from the input and white Gaussian noise, as before. Importantly, we used a single global noise scale, $\sigma$, instead of having a separate, tunable noise parameter for each cell, $\sigma_i$, because the network was able to use its recurrent synaptic connectivity to modulate each cell's variability. Moreover, as the same recurrent weights also allowed the responses to be potentially correlated, we made the learning task for the network more challenging by fixing the reconstruction coefficients such that the corresponding posteriors now showed correlations (Fig. 4B and D, thin lines; cf. Fig. 1E, H). As the reconstruction coefficients were fixed, for simplicity, we replaced the sparseness cost (proportional to the absolute value of responses, Eq. 2) with a simple squared cost on activities (corresponding to a Gaussian prior). We optimised all other network parameters (including both feed-forward and recurrent weights) as before, and incorporated the obvious extension of the reliability cost used previously: reliability cost $\propto 1/\sigma^2$ (Fig. 1B).

After optimisation, neural activities changed dynamically (while keeping the input fixed in a trial; Fig. 4C, red trace) such that the stationary distribution of responses for each input (obtained by collecting responses within a trial) closely matched the required posterior distribution (Fig. 4C, gray scale map, and D-F). Remarkably, the network not only achieved the required mean and standard deviation for the activities of single neurons (Fig. 4E-F, red and blue), but the correlations among cells were also well matched to the posterior correlations (Fig. 4E-F, green). Thus, posterior inference over complex, correlated posteriors has re-emerged, despite the simplicity of the reliability cost, which now depended on only a single parameter of the neural network, $\sigma$. To understand why this occurred, we note that the reliability cost can be rewritten as (Methods Sec. 3.2.2),

$$\text{reliability cost} \propto \sum_i \frac{1}{\lambda_i^2} \qquad (9)$$

where $\lambda_i^2$ represents effective (stationary) variability in the *population*, that is the variance along each principle component (formally, the eigenvalues of the stationary covariance), rather than merely the variability in the response of *individual* neurons. This population quantity bounds (and approximates) the entropy of the *population* response in the same way that the inverse variance of individual cells bounded the entropy of single neuron responses in the feed-forward case (SI Fig. 6). Through the link to variational inference we described above (Eq. 6), this guarantees that response distributions will match the corresponding posteriors – but now also in the presence of correlations and even without directly tunable neuron-specific variability.

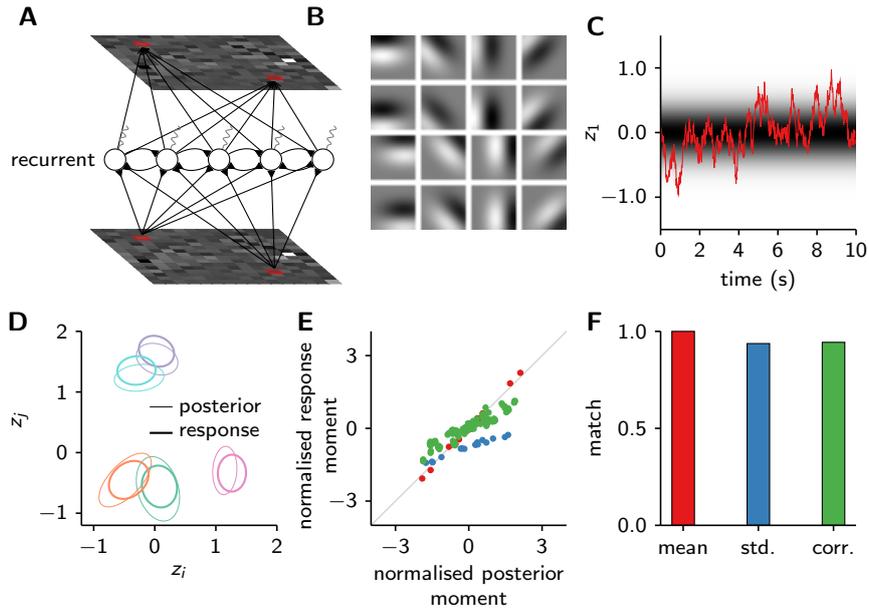

**Figure 4:** A recurrent neural network sampler. **A.** Schematic diagram of the network, with additional recurrent connections (cf. Fig. 1A). **B.** Reconstruction coefficients (fixed, not learned), forming edge detecting (Gabor) filters as in Ref. [6]. **C.** Dynamics of a neuron over 10 s. Gray scale map displays the underlying stationary distribution. **D.** Covariance ellipses displaying example neural response distributions (thick lines) and the corresponding true posteriors (thin lines) for a representative input. Different colors correspond to different pairs of cells. **E.** Mean (red), standard deviation (blue), and correlation (green) of the posterior (x-axis) and response distributions (y-axis), normalized by the standard deviation of the relevant posterior moment. Different dots correspond to different cells (cell pairs). **F.** Match between posterior and response distribution moments across images (colors as in E).

## 1.6 Recurrent networks optimized for sampling speed

If variability in neural networks can be used to sample from posterior distributions, as we have shown, then for this to be useful for approximate inference network dynamics need to be fast in order to provide a large number of (statistically independent) samples over behaviorally relevant (and thus usually limited) time scales. It has been suggested that oscillatory and excitatory-inhibitory dynamics in the cortex might be particularly useful for such fast sampling [53, 54, 52]. To achieve fast sampling, we revisited the simple recurrent neural network studied above (Fig. 4, Eq. 8), and incorporated an additional term into the cost function (Eq. 6a) that measured the accuracy of an estimate of the mean of the distribution, based on samples taken over a finite period of time. Incorporating this additional term in the objective did not perturb the learned stationary distributions (compare Fig. 4D-F with Fig. 5A-C). However, the dynamics changed dramatically (Fig. 5D,F), with the fast variant (right) moving across the stationary distribution (gray scale map) more rapidly than the original, slow variant (left). We quantified these effects by the average autocorrelation (Fig. 5E), which decayed slowly ($\sim$100 ms) for the slow model, and rapidly ($\sim$20 ms) for the fast model. As a result, the mean of the distribution was indeed estimated more rapidly using the fast than the slow variant (Fig. 5G).

## 2 Discussion

We have shown that neural circuits optimised for performance in standard unsupervised learning tasks under basic biophysical constraints on their reliability develop specific patterns of response variability. The distributions of responses that emerge in these networks closely match the posterior distributions inferred under a well-calibrated probabilistic model of their inputs. This implies that the responses of these networks represent samples from the required posterior distributions, thus "automatically" implementing a state-of-the-art probabilistic inference algorithm: sampling-based variational inference [24, 25, 55]. Critically, unlike classical machine learning algorithms, these networks do not optimize parameters for each data-item individually; instead they must use the same set of synaptic weights to map every input onto an appropriate approximate posterior, an approach known as "amortised inference" in machine learning that has been recently shown to improve the efficiency of inference and learning in large (and deep) neural networks [24, 25, 55].

Recent work by [56] considered another approach to train neural networks to perform Bayesian inference without using an explicit hand-crafted inference algorithm. In particular, they considered tasks whose optimal solution required Bayesian inference and then learned the tasks using highly flexible supervised neural networks. Thus, when their network attained near-optimal performance, it must have learned to perform Bayesian inference, at least implicitly. Indeed, the optimised networks represented uncertainty in mean neuronal responses (and in particular in their across-population sparseness), rather than their variability. There are two critical differences between their approach and ours. First, in their approach, learning to perform Bayesian inference required a supervision signal, and while this is present in the small-scale cognitive or psychophysical tasks they considered, in natural environments feedback is often weak or non-existent [57]. In contrast, in complex natural environments without feedback it is necessary to use unsupervised approaches, such as ours, to perform complex inferences concerning the structure of natural stimuli. Second, they considered deterministic networks, thus ignoring the biophysical cost of reliability and the consequences of its interplay with performance criteria. Consequently, it is unknown how response distributions would be shaped by performance-variability trade-offs in a stochastic generalisation of these models, as these deterministic networks are by construction unable to account for structured variability in cortical responses [58, 47].

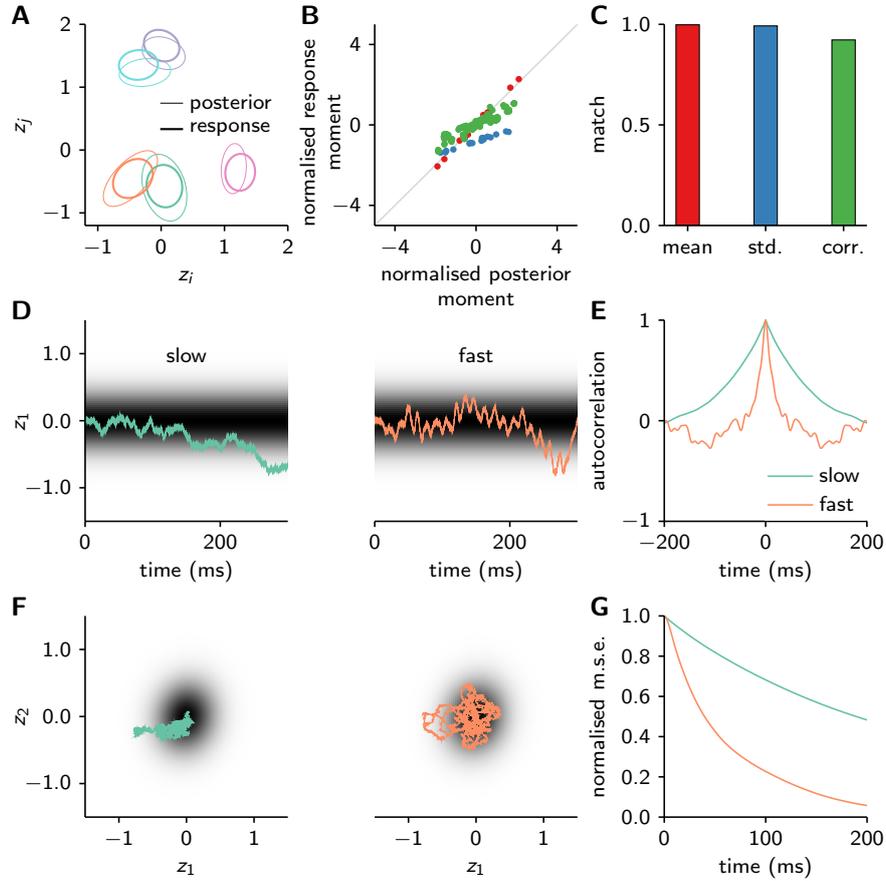

**Figure 5:** Fast sampling in recurrent neural networks. **A.** Covariance ellipses displaying example neural response distributions (thick lines) and the corresponding true posteriors (thin lines) for a representative input. Different colors correspond to different pairs of cells (cf. Fig. 4D). **B.** Mean (red), standard deviation (blue), and correlation (green) of the posterior (x-axis) and response distributions (y-axis), normalized by the standard deviation of the relevant posterior moment. Different dots correspond to different cells (cell pairs, cf. Fig. 4E). **C.** Match between posterior and response distribution moments across images (colors as in B, cf. to Fig. 4F). **D.** The activity of a representative neuron (colored traces) in the slow (left, blue; cf. 4) and the fast network (right, orange). Gray scale map shows the underlying posterior distribution. **E.** The average autocorrelation for neurons in the slow (blue) and fast network (orange). **F.** Joint activity of two representative neurons (colored traces) in the slow (left, blue) and the fast network (right, orange). Gray scale map shows the underlying posterior distribution. Both traces show activity over a period of 300 ms. Note that the fast network covers the full extent of the posterior over this period, while the slow network only visits a small fraction of it. **G.** Normalized mean squared error (MSE) as a function of the time period over which neural activities produced by the slow (blue) or fast network (orange) are averaged. The rapid decay of the MSE in the fast network is partly due to negative autocorrelations at intermediate time scales cancelling large positive autcorrelations at the shortest time scales (panel E).

Our work was a first step towards understanding how biophysical energetic costs shape not only mean neural responses but also variability, and as such, it relied on a variety of simplifications. First, we considered a simple, inverse-square form for the cost of variability, which nevertheless emerged in a broad range of feedforward and recurrent models, and thus provided a reasonable initial approximation. Second, we used a variety of simplifications in our neural networks. Most importantly, in the recurrent networks, neurons had linear activation functions (but see the feedforward deep network for neurons with nonlinear activation functions). While the analytic derivations do apply more broadly, such as in cases where the cost of reliability is any (negative) power law, we also expect the essential intuition — that biophysical costs on reliability encourage neural responses to sample the posterior — to hold in more complex and biologically realistic settings where the analytic proofs no longer hold.

We studied the consequences of performance optimisation under biological constraints while remaining agnostic about the mechanism by which it is achieved. To carry out optimisation in our simulations, we used a generic stochastic gradient descent algorithm [59], and while it may be approximately implemented by synaptic plasticity [31, 32, 33], real neural circuits may use a variety of other learning rules to optimise their performance [28, 60].

There is a considerable body of neural data suggesting that, as in the networks studied here, neural responses sample posterior distributions. Such a sampling-based representation of uncertainty has been shown to account for patterns of neural variability in the visual cortex [20] and their modulation by feedforward visual inputs [21] as well as feed-back task-related signals [61]. However, previous approaches to understanding how these patterns might have emerged via neural circuit dynamics employed a top-down approach which began by choosing an inference algorithm from the machine learning literature and then suggested an implementation in neural hardware [62, 63, 52]. While this was sometimes effective, the relevant machine learning methods always required strong constraints that messy, complex biological networks are unlikely to be able to satisfy, and thus strong approximations were usually required. In contrast, we took a bottom-up approach, showing that posterior sampling emerges naturally as a consequence of the trade-off between minimizing reconstruction error, and minimizing the energetic cost of achieving highly reliable responses.

The natural emergence of neural response distributions resembling posterior sampling does raise a question about how to interpret corresponding neural data. In particular, this data has been taken to suggest that this representation is there to support a specific computational strategy in the brain, sampling-based probabilistic inference. However, if, as shown here, these posterior-like response distributions emerge naturally as a consequence of an accuracy-reliability tradeoff then they are not necessarily evidence of this specific computational strategy: the same response distributions would emerge even if downstream circuits simply used mean activity, and ignored information about uncertainty embodied in variability. To understand whether these patterns are thus an "epiphenomenon" emerging from our tradeoff, or are evidence of probabilistic computation in the brain, we will need to understand both whether the brain expends additional energy in optimizing the sampling process (such as speeding it up), and to what extent downstream circuits exploit variability to extract information about uncertainty.

# 3 Methods

## 3.1 Variational inference

We start by giving a brief introduction to variational inference. A standard way to learn a probabilistic internal model of inputs is to maximise the marginal likelihood of the model, i.e. to find the setting of the parameters with which the internal model predicts experienced inputs with the highest probability, integrating over all possible settings of the latents. A fundamental observation is that it is possible to bound the marginal likelihood using the variational free-energy [22],

$$-\log P_\theta(x) \leq \mathcal{F} = -\log P_\theta(x) + D_{KL}\left(Q_\psi(z|x) \,||\, P_\theta(z|x)\right). \tag{10}$$

where $\mathcal{F}$ is the free energy, $P_\theta(x)$ is the marginal likelihood, $P_\theta(z|x)$ is the (exact) posterior over latent variables $z$ given an input $x$ under the current internal model, whose parameters are $\theta$ (e.g. the reconstruction coefficients, $\mathbf{A}$, in Fig. 1), $Q_\psi(z|x)$ is an approximation to this posterior (our response distribution) parametrised by a separate set of parameters $\psi$ (e.g. the synaptic weights, $\mathbf{W}$, and single-neuron variabilities, $\sigma_i$, in Fig. 1), and $D_{KL}$ denotes the Kullback-Leibler (KL) divergence. This formulation reveals that minimizing the variational free energy will encourage the KL term to go to zero, and hence encourage the response distribution, $Q_\psi(z|x)$, to approach the true posterior, $P_\theta(z|x)$. For a sufficiently flexible response distribution, the KL divergence will go to zero, at which point the response distribution and true posterior become equal, and the objective becomes equal to $\log P_\theta(\mathbf{x})$: so the problem becomes one of maximizing the marginal likelihood w.r.t. the generative parameters, $\theta$. Critically, the free energy can also be rewritten in a different form [64] that is closely related to the cost function we used to train our networks:

$$\mathcal{F} = -E_{Q_\psi(z|x)}\left[\log P_\theta(x, z)\right] - H\left[Q_\psi(z|x)\right], \tag{11}$$

where $E_{Q_\psi(z|x)}\left[\log P_\theta(x, z)\right]$ is a measure of how well the settings of latent variables sampled from the response distribution explain (or encode) the observations, and $H\left[Q_\psi(z|x)\right]$ is the entropy of the response distribution, which is an information-theoretic measure of the overall variability of the sampled activities.

### 3.2 Relating the biological reliability cost to variational inference

To understand how biological systems might be approximating variational inference, we noted that the entropy term in the variational free energy penalises highly reliable responses, and asked how such a cost might arise in a biological system. In particular, in this section, we relate univariate and multivariate Gaussian entropies to the biologically motivated inverse-square power-law form for the cost of reliability, and in the subsequent section (Sec. 3.3), we give additional biophysical models under which this inverse-square law form emerges naturally.

#### 3.2.1 Univariate/uncorrelated case

As discussed above, variational inference penalises reliability using a cost based on the entropy. When the response distribution, $Q_\psi(\mathbf{z}|\mathbf{x})$, is an independent Gaussian distribution (as in our feedforward networks), the entropy becomes a sum of separate terms, $H_i$, each denoting the entropy of the corresponding Gaussian latent variable with variance $\sigma_i^2$:

$$-H_i = -\tfrac{1}{2}\log \sigma_i^2 - \tfrac{1}{2}\log(2\pi e). \tag{12}$$

To relate the entropy to the inverse-square power-law form for the reliability cost in the main text (Eq. 3), we begin by introducing a scale parameter, $s$, that is the same across cells,

$$-H_i = \tfrac{1}{2}\log\left(\frac{s}{\sigma_i}\right)^2 - \tfrac{1}{2}\log(2\pi e s^2). \tag{13}$$

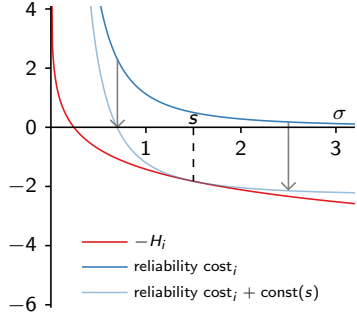

**Figure 6:** Diagram showing how the reliability cost (with a constant offset) approximates and bounds $-H_i$.

Then, we approximate (and bound) the logarithm (a concave function) by its first-order Taylor expansion at 1, i.e. $\log x \approx x - 1$ (and $\log x \leq x - 1$), and recover the reliability cost that we derived using biophysical considerations (Eq. 3, Fig. 6):

$$-H_i \leq \frac{1}{2}\left(\left(\frac{s}{\sigma_i}\right)^2 - 1\right) - \tfrac{1}{2}\log(2\pi e s^2) = \text{reliability cost}_i + \text{const}(s), \tag{14}$$

where,

$$\text{reliability cost}_i = \frac{s^2}{2\sigma_i^2}, \tag{15}$$

$$\text{const}(s) = \tfrac{1}{2}\log 2\pi e^2 s^2. \tag{16}$$

Notably, this approximation has one parameter, $s$. In principle, $s$ is set by the underlying biology, as it determines the relative tradeoff between the reliability cost and the sparse coding cost (or in general the log-joint probability). Instead of adjusting $s$ manually, we instead set $s$ by incorporating it in our optimization procedure, by finding the setting for $s$ that optimized the bound in Eq. (14) (see also section 3.4.1).

### 3.2.2 Multivariate/correlated case

When the response distribution, $Q_\psi(\mathbf{z}|\mathbf{x})$, is a correlated Gaussian distribution (as in our recurrent networks), its entropy is given by

$$-H = -\frac{1}{2}\sum_i \log \lambda_i^2 - \frac{N}{2}\log 2\pi e = \tfrac{1}{2}\sum_i \log \frac{s^2}{\lambda_i^2} - \frac{N}{2}\log 2\pi e s^2 \tag{17}$$

where $\lambda_i^2$ are the eigenvalues of the covariance matrix, $\Sigma$ (i.e. the variances along the principle component directions). As before, we approximate (and bound) the logarithm using a first-order Taylor expansion centred at 1, i.e. $\log x \approx x - 1$ (and $\log x \leq x - 1$),

$$-H \leq \frac{1}{2}\sum_i \left(\left(\frac{s}{\lambda_i}\right)^2 - 1\right) - \frac{N}{2}\log 2\pi e s^2. \tag{18}$$

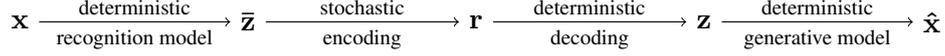

**Figure 7:** Describing the paradigm under which the recognition model deterministically transforms data, $\mathbf{x}$, into mean latents, $\bar{\mathbf{z}}$, which are stochastically encoded in neural activities, $\mathbf{r}$, and can then be deterministically read-off, to give the actual settings for the latent variables represented by the system, $\mathbf{z}$, which can then be used to reconstruct the input, $\hat{\mathbf{x}}$.

which gives us almost exactly the expression we had in the univariate case, except that the variance for individual latent variables, $\sigma_i^2$ has been replaced by the variance along the correlated, principle component directions, $\lambda_i^2$. In matrix form, this is equivalent to,

$$-H \leq \tfrac{s^2}{2} \text{Tr}\left(\mathbf{\Sigma}^{-1}\right) + N \ \text{const}(s). \tag{19}$$

While we have already connected the univariate entropy to biophysical energetic costs, to do the same for the multivariate entropy, we need to consider the additional biophysical models given in the subsequent section.

### 3.3 Cost of reliability in additional biophysical models

In the previous section, we showed that our inverse-square form for the reliability cost forms a useful approximation to and bound on the entropy, and thus gives rise to approximate variational inference. In this section we consider biophysical models under which this inverse-square reliability cost might emerge, in addition to the suggestion in the main text of having a small population of stochastic elements (e.g. cells) encode each latent variable.

These models are premised on the idea that the data, $\mathbf{x}$, deterministically gives rise to an implicit "mean" latent variable, $\bar{\mathbf{z}}$, which is then stochastically encoded in neural responses, $\mathbf{r}$. The biophysical reliability cost relates to this stochastic encoding step, and is larger for a more reliable encoding of the underlying mean latent, $\bar{\mathbf{z}}$. As $\bar{\mathbf{z}}$ is never explicitly represented in the neural system, downstream circuits can only read out the stochastic variable, $\mathbf{r}$, and the best they can do is to deterministically decode the latent variables given those responses, $\mathbf{z} = \mathbf{z}(\mathbf{r})$ (Fig. 7). Thus, we can translate a biophysical reliability cost attached to the neural responses, $\mathbf{r}$, to a reliability cost for the decoded latent variables, $\mathbf{z} = \mathbf{z}(\mathbf{r})$.

#### 3.3.1 Representing a single latent variable using Poisson spikes

Here we consider the reliability cost for a Poisson neuron. In particular, our stochastic latent variable, $z_i$, is a deterministic decoding of neural responses, $r_i$, which are Poisson-distributed with a rate based on the deterministic mean value for the latent variable, $\bar{z}_i$,

$$r_i \sim \text{Poisson}\left(g_i \, \bar{z}_i(x)\right), \tag{20}$$

where the Poisson rate is given by the product of the latent variable with a gain factor $g_i$. Downstream circuits can estimate the value of $z_i$ that gave rise to the neural response using,

$$z_i = r_i/g_i. \tag{21}$$

Thus, the mean of the estimate is unbiased,

$$\mathrm{E}\left[z_i|\mathbf{x}\right] = \mathrm{E}\left[z_i|\bar{z}_i(\mathbf{x})\right] = \bar{z}_i(\mathbf{x}), \tag{22}$$

and its variance falls with increasing gain $g_i$ (and hence with increasing average firing rate),

$$\mathrm{Var}\left[z_i|\mathbf{x}\right] = \mathrm{Var}\left[z_i|\bar{z}_i(\mathbf{x})\right] = \frac{\bar{z}_i(\mathbf{x})}{g_i}. \tag{23}$$

Thus, the only way to improve reliability for a fixed $\bar{z}_i$ is to increase the gain, and hence to increase the expected firing rate. If we take the cost to be proportional to the expected firing rate, we recover the inverse power law form for the reliability cost used in the main text,

$$\text{reliability cost}_i \propto \mathrm{E}\left[r_i|\mathbf{x}\right] \propto g_i \propto \frac{1}{\mathrm{Var}\left[z_i|\mathbf{x}\right]}. \tag{24}$$

### 3.3.2 Linear embedding of latent variables in a larger population of neural responses

Previously, we assumed that each cell encodes only one latent feature, $z_i$, either alone or as part of a small population. More realistically, cells in the brain respond to multiple features in the environment, and thus jointly encode multiple quantities. Suppose $n_z$ latents are encoded in the responses of $N$ neurons ($N > n_z$) using a linear embedding,

$$\mathbf{r} = \mathbf{W}\,\bar{\mathbf{z}}(\mathbf{x}) + \sigma\,\boldsymbol{\eta}, \tag{25}$$

where $\boldsymbol{\eta}$ is independent, unit-variance noise. In the Poisson case described above, the variability of $\mathbf{r}$ depended on the mean via the Poisson assumption, whereas here we fixed the variability for simplicity. In particular we take the biophysical energetic cost of reliability to be,

$$\text{reliability cost} \propto \frac{N}{\sigma^2}, \tag{26}$$

and our goal is to write this cost as a function of the implied noise over the decoded latents, $\mathbf{z}$ (i.e. the covariance $\boldsymbol{\Sigma}$), rather than in terms of the noise in the neural responses. Before doing this, however, we needed to constrain our system, because it is now possible to reduce the effective noise to zero, and achieve a perfect representation by increasing the magnitude of the weights, $\mathbf{W}$, whilst keeping the noise, $\sigma$ fixed. In particular, we fixed the scale of the weights to

$$\langle W_{ij}^2 \rangle \equiv \frac{1}{Nn_z} \sum_{ij} W_{ij}^2 \propto 1/n_z, \tag{27}$$

which further ensures that changes in $r_i$ induced by changes in $\mathbf{z}$ have constant variance, even as the number of latents increases.

To express the reliability cost in terms of variability in the decoded variables, $\mathbf{z}$, we begin by considering the ideal decoder itself,

$$\mathbf{z} = \left(\mathbf{W}^T\mathbf{W}\right)^{-1}\mathbf{W}^T\mathbf{r}. \tag{28}$$

again implying an unbiased mean estimate,

$$\mathrm{E}\left[\mathbf{z}|\mathbf{x}\right] = \mathrm{E}\left[\mathbf{z}|\bar{\mathbf{z}}(\mathbf{x})\right] = \bar{\mathbf{z}}(\mathbf{x}). \tag{29}$$

Thus, the variability in the decoded $\mathbf{z}$ is given by,

$$\boldsymbol{\Sigma} = \text{Cov}\left[\mathbf{z}|\mathbf{x}\right] = \left(\mathbf{W}^T\mathbf{W}\right)^{-1}\mathbf{W}^T\left(\sigma^2\mathbf{I}\right)\mathbf{W}\left(\mathbf{W}^T\mathbf{W}\right)^{-T} = \sigma^2\left(\mathbf{W}^T\mathbf{W}\right)^{-1}. \tag{30}$$

To recover the reliability cost, we take the trace of the inverse covariance,

$$\text{Tr}\left(\boldsymbol{\Sigma}^{-1}\right) = \frac{\text{Tr}\left(\mathbf{W}^T\mathbf{W}\right)}{\sigma^2} \propto \frac{N}{\sigma^2} \propto \text{reliability cost}. \tag{31}$$

where the first proportionality comes from the constraint Eq. (27), combined with $\text{Tr}\left(\mathbf{W}^T\mathbf{W}\right) = \sum_{ij} W_{ij}^2$. Thus, the reliability cost is (up to an additive constant) proportional to our usual approximation to and bound on the entropy (Eq. 19).

### 3.3.3 Linear recurrent network

Here we consider the reliability cost for the dynamics of a linear recurrent network, with dynamics that potentially depend in the input (which we will exploit later). As in the previous section, we take the $\sigma_i$ to be fixed to the same value across cells, $\sigma_i = \sigma$, and the reliability cost to be given by Eq. (26). The key difference is that here we assume each neuron corresponds to one latent, $\mathbf{r} = \mathbf{z}$, with these responses being governed by dynamics,

$$\dot{\mathbf{r}} = -\frac{1}{\tau}\boldsymbol{\Pi}\,\mathbf{r} + \frac{\sigma}{\sqrt{\tau}}\,\boldsymbol{\eta}, \tag{32}$$

where we neglect any additive $\mathbf{x}$-dependent input as it does not affect response variability. In the simplest case (Eq. 8), we have,

$$\boldsymbol{\Pi} = \mathbf{I} - \mathbf{W}^{\text{rec}}. \tag{33}$$

Again, our aim is to write the cost of reliability in terms of the covariance of the trivially decoded $\mathbf{z} = \mathbf{r}$, rather than the input noise. To do so, we write the dynamics matrix as,

$$\tfrac{1}{\tau}\boldsymbol{\Pi} = \left(\tfrac{\sigma^2}{2\tau}\mathbf{I} + \mathbf{S}(\mathbf{x})\right)\boldsymbol{\Sigma}^{-1}(\mathbf{x}), \tag{34}$$

where $\sigma^2/\tau\mathbf{I}$, is the covariance of the process noise in Eq. (32), $\boldsymbol{\Sigma}(\mathbf{x})$ is the stationary covariance, and $\mathbf{S}(\mathbf{x})$ is a skew-symmetric matrix [53]. Taking the trace of both sides of Eq. (34), the expression simplifies considerably as the skew-symmetric matrix is eliminated, because the trace of the product of a symmetric and skew-symmetric matrix is zero, and the trace of $\boldsymbol{\Pi}$ is $N$ as there are no self-connections,

$$\frac{N}{\tau} = \frac{\sigma^2}{2\tau}\text{Tr}\left(\boldsymbol{\Sigma}^{-1}\right) \tag{35}$$

thus, we can write the reliability cost as,

$$\text{reliability cost} \propto \frac{N}{\sigma^2} = \frac{1}{2}\text{Tr}\left(\boldsymbol{\Sigma}^{-1}\right). \tag{36}$$

Thus, the reliability cost (up to an additive constant) is proportional to our usual approximation to and bound on the entropy (Eq. 19).

## 3.4 Robustness with respect to parameters of the biological cost

Here, we study the expected robustness of our results to changes in the scaling factor $s$, which controls the relative importance of the reliability cost and sparse coding cost (and is in principle set by the actual biophysical cost of achieving some level of reliability). We also investigate the effect of the exponent in the power-law form of the reliability cost that we have proposed.

To understand how the approximate posterior variance, $\sigma^2$ (omitting the cell index subscript for brevity) changes with these parameters, we generalise the derivation above, defining a family of approximations to the entropy,

$$H_p = -\tfrac{1}{p}\left(\left(\tfrac{s}{\sigma}\right)^p - 1\right) + \tfrac{1}{2}\log 2\pi e s^2 \tag{37}$$

such that reliability cost + const $= -H_{p=2}$. Notably, all these approximations (i.e. for any positive exponent $p$ are valid lower-bounds on the true entropy, $H_p \leq H$.

To understand the robustness (with respect to $p$ and $s$) of inferences under this generalised reliability cost, we combine it with the simplest possible Bayesian posterior distribution, i.e. a Gaussian posterior over a 1-dimensional latent variable. Recall

$$-\mathcal{F}_p = \mathcal{L}_p = \mathrm{E}_{\mathrm{Q}(z|x)}\left[\log \mathrm{P}(z|x)\right] + H_p + \mathrm{const}, \tag{38}$$

where the constant is $\log \mathrm{P}(x)$. Taking the true posterior to be Gaussian, with mean $m$ and variance $v^2$, this becomes,

$$\mathcal{L}_p = \mathrm{E}_{\mathrm{Q}(z|x)}\left[-\frac{(z-m)^2}{2v^2}\right] - \frac{s^p}{p\sigma^p} + \mathrm{const}, \tag{39}$$

where the constant now also incorporates the normalizing constant from the true posterior and terms from the bound on the entropy. Evaluating the expectation, assuming that the approximate posterior is Gaussian with matched mean $m$ and potentially unmatched variance, $\sigma^2$, gives,

$$\mathcal{L}_p = -\frac{\sigma^2}{2v^2} - \frac{s^p}{p\sigma^p} + \mathrm{const}. \tag{40}$$

To find the maximum of the objective w.r.t. $\sigma$, we set the gradient to zero,

$$0 = \frac{\partial \mathcal{L}_p}{\partial \sigma} = -\frac{\sigma}{v^2} + \frac{s^p}{\sigma^{p+1}}. \tag{41}$$

and solve for $\sigma$,

$$\sigma = s^{p/(p+2)} v^{2/(p+2)} = \sqrt[p+2]{s^p v^2}. \tag{42}$$

We can think of this as a "weighted geometric average", between the true posterior standard deviation, $v$, and the scale parameter, $s$. We therefore expect our biologically plausible objectives to shrink all the true posterior standard deviations towards $s$, to a degree that depends on the power-law, $p$. In particular, as $p$ approaches zero the weight given to the true posterior standard deviation, $v$, increases, until at $p = 0$, we have $\lim_{p \to 0} \sigma = v$. For our usual power law with $p = 2$, we get the geometric average between the true posterior standard deviation, $v$, and the scale, $s$,

$$\sigma(p=2) = \sqrt{sv} = A^{1/4} v^{1/2}, \tag{43}$$

where, in this generalised setting, we have taken,

$$A = s^p. \tag{44}$$

The resulting effects are displayed in Fig. 8. For illustration purposes, we consider a case where the true posterior is Gaussian over two independent latent variables, such that our analytical derivations apply independently to each (with the same $p$ and $A$, or equivalently, the same $s$). As predicted by the above arguments, the approximate posteriors end up more spherical than the true posterior, because the standard deviation along both directions moves towards a common value of $s$ (Fig. 8A). This "sphericalising" pattern is emphasised by Fig. 8B, which shows that as the true posterior standard deviation increases, the approximate posterior standard deviation also increases, but to a lesser degree (note different coloured lines correspond to the settings of $A$ in Fig. 8A). Second, we see that while $A$ does influence the approximate posterior standard deviation (Fig. 8AB), the influence is relatively small: the $1/4$ power in Eq.(43) suppresses changes dramatically. This is evident in Fig. 8A where $A$ changes by a factor of 4, and while this does change the scale of the approximate posterior, the changes are remarkably small. This is particularly evident in Fig. 8C, where we see that the approximate posterior standard deviation changes slowly as $A$ changes. Such a weak dependence of the resulting posterior distribution on $A$ shows that $A$ needs not be fine-tuned; this is reassuring, as $A$ trades off the relative importance of the reconstruction error and biophysical energetic costs, and thus $A$ is set by the underlying biology, and is unlikely to be set to the optimal value for Bayesian inference. Third, we see that the power-law exponent, $p$, controls the quality of the approximation, with $p = 0$ being perfect, with increasingly approximate distributions as $p$ increases. This is evident in examples (Fig. 8D), and more systematically in the relationship between the true and approximate standard deviations, which become identical as $p$ approaches 0 (Fig. 8EF).

### 3.4.1 Optimizing $s$

In Fig. 8DEF, we set $s$ to its optimal value, which depends on $p$, $v_1$ and $v_2$ (i.e. the true posterior standard deviations). In our toy example, we can in fact derive the optimal $s$ analytically and garner additional insight into the function of this parameter.

Naively, we might try to optimize $s$ in the univariate model above, by setting the gradient to zero,

$$0 = \frac{\partial \mathcal{L}_p}{\partial s} = \frac{s^{p-1}}{\sigma^p} + \frac{1}{s}, \tag{45}$$

and solving for $s$,

$$s = \sigma, \tag{46}$$

thus, the biologically motivated approximation to the entropy reduces to the true entropy for any $p$,

$$H_p(s = \sigma) = \tfrac{1}{2} \log 2\pi e s^2 = H. \tag{47}$$

This approach broke down because we considered only a single cell, so instead, we need to consider the optimal $s$ when we have multiple cells. We begin by writing down the $\mathcal{L}_p$ objective for multiple cells, then substitute optimized $\sigma_i$, and optimize the final, univariate objective with respect to $s$. The objective is,

$$\mathcal{L}_p = -\sum_i \frac{\sigma_i^2}{2v_i^2} - \frac{s^p}{p\sigma_i^p} + \tfrac{N}{2} \log 2\pi e s^2 + \text{const.} \tag{48}$$

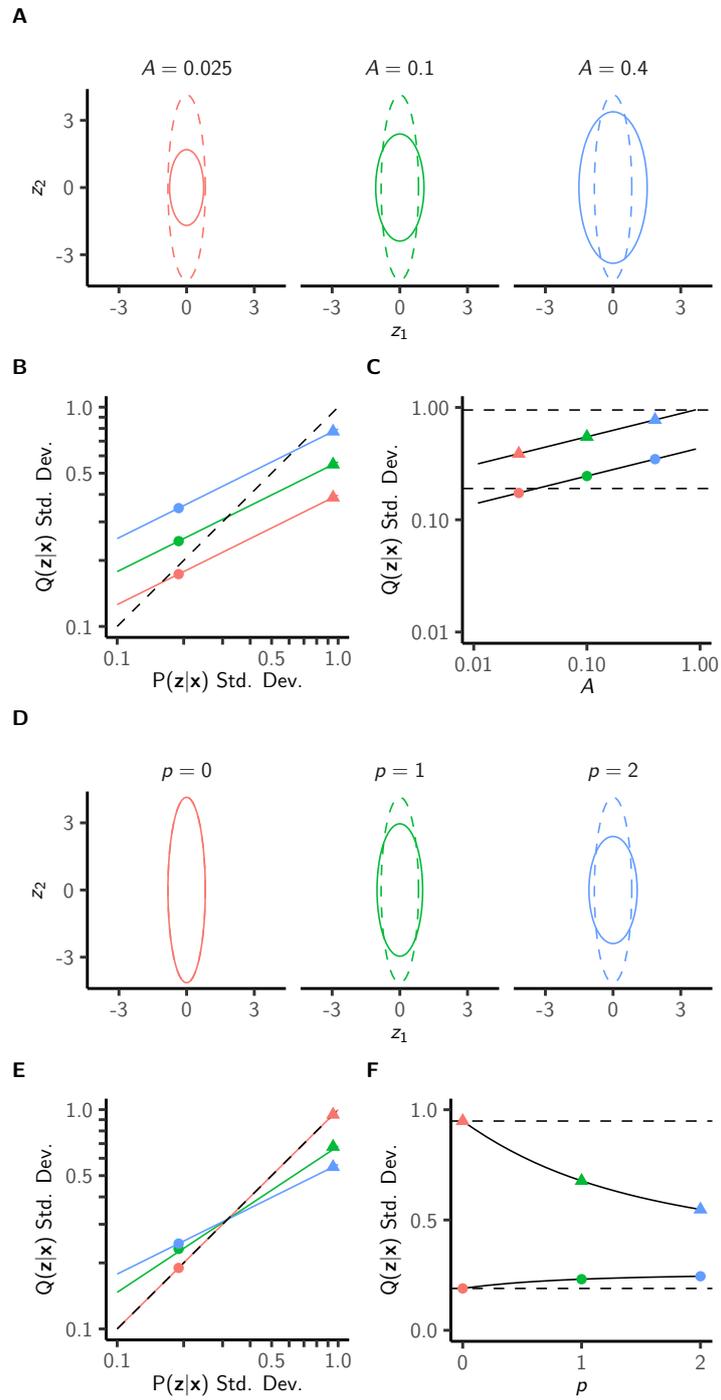

**Figure 8:** **A.** True (dashed) and approximate (solid) posteriors for three different settings of $A$, the multiplicative constant that scales the reliability cost. The three values of $A$ shown are optimal (see Sec. 3.2.1 and Sec. 3.4.1 $A = 0.1$, center), a factor of 4 smaller than optimal ($A = 0.025$, left), and a factor of 4 larger than optimal ($A = 0.4$, right). While the overall size of the approximate posterior changes, the quality of the approximation appears to be fairly consistent, in the sense that the longer axis of the true posterior is always the longer axis of the approximate posterior. **B.** The relationship between the true posterior and approximate posterior for the three different settings of $A$ displayed in **A**. The circular points represent the actual values along the $z_1$ axis displayed in **A**, whereas the triangular points represent the values along the $z_2$ axis. The dashed line is the identity line, representing the ideal. Notably, there is a general bias for the approximate posterior to be smaller than the true posterior, that emerges from standard properties of the variational distribution [65, 66]. **C.** The approximate posterior standard deviations for $z_1$ (lower) and $z_2$ (upper) displayed in **A**. The lower and upper dashed lines represent the two true standard deviations. Notably, the approximate standard deviation is remarkably insensitive to changes in $A$: to achieve a factor-of-two change in the posterior standard deviation, we need to change $A$ by a factor of $2^4 = 16$. **DEF.** as in **ABC** respectively, but where we change the power $p$, rather than $A$. Note that as $p$ decreases towards 0, the approximation becomes perfect.

substituting optimized $\sigma_i$ from Eq. (42) and simplifying,

$$\mathcal{L}_p = -\frac{1}{q}\sum_i \left(\frac{s}{v_i}\right)^q + \tfrac{N}{2}\log 2\pi e s^2 + \text{const.} \tag{49}$$

where,

$$q = 2p/(p+2). \tag{50}$$

To optimize, we set the gradient to zero,

$$0 = \frac{\partial}{\partial s}\mathcal{L}_p = -s^{q-1}\sum_i v^{-q} + Ns^{-1}. \tag{51}$$

Finally, solving for $s$ gives,

$$s = \left[\tfrac{1}{N}\sum_i v_i^{-q}\right]^{-1/q}, \tag{52}$$

Which is the generalised mean with power $-q$. For our usual case ($p = 2$), this is the Harmonic mean,

$$s(p=2) = \left(\tfrac{1}{N}\sum_i v_i^{-1}\right)^{-1}. \tag{53}$$

Thus, the optimal scale, $s$, should be set to some central estimate of the true posterior standard deviations, $v_i$.

## 3.5 Optimizing the free energy

We optimized the free energy (Eq. 11) using stochastic gradient descent, which required us to be able to differentiate the free energy with respect to the generative and recognition parameters. Optimizing with respect to the generative parameters, $\theta$, is straightforward, and leads to parameter updates of the form $\theta \leftarrow \theta - \alpha \Delta \theta$, where $\alpha$ is a learning rate, $\Delta \theta$ is a sample-based (Monte-Carlo) estimate of the gradient

$$\nabla_\theta \mathcal{F} = -\mathrm{E}_{\mathrm{Q}_\psi(z|x)} \left[ \nabla_\theta \log \mathrm{P}_\theta(x, z) \right], \tag{54}$$

and where $\nabla_\theta$ is the gradient operator w.r.t. $\theta$. It is more difficult to optimize with respect to the recognition parameters, $\psi$, as they control the distribution over which the expectation is taken. To solve this problem, we use the reparameterisation trick [24, 25]: we parameterize samples from the recognition model with i.i.d. random variables, $\epsilon$, transformed by potentially complex but smooth functions $z(\epsilon; \psi)$ (e.g. using the inverse cumulative density function transformation). Consequently, the variational free energy can be rewritten as an expectation over these i.i.d. random variables,

$$\mathcal{F} = -\mathrm{E}_\epsilon \left[ \log \mathrm{P}_\theta(x, z(\epsilon; \psi)) \right] - \mathrm{H}\left[ \mathrm{Q}_\psi(z|x) \right]. \tag{55}$$

This makes it easy to optimize w.r.t. the recognition parameters by computing the corresponding gradient,

$$\nabla_\psi \mathcal{F} = -\mathrm{E}_\epsilon \left[ \nabla_\psi \log \mathrm{P}_\theta(x, z(\epsilon; \psi)) \right] - \nabla_\psi \mathrm{H}\left[ \mathrm{Q}_\psi(z|x) \right], \tag{56}$$

where the updates for $\psi$ are given by analogy with those for $\theta$.

## 3.6 Feed-forward neural networks

For our data, $\mathbf{x}$, we use binarised MNIST images. The generative model begins by drawing top-layer activations IID from a Gaussian,

$$\mathrm{P}(\mathbf{z}_N) = \mathcal{N}\left(\mathbf{z}_N; \mathbf{0}, \mathbf{D}_N^{\mathrm{gen}}\right) \tag{57}$$

where $\mathbf{D}_i$ is a diagonal matrix of free parameters. At each subsequent layer, these activations are propagated through a non-linear function $\mathbf{f}_i$, and Gaussian noise is added,

$$\mathrm{P}(\mathbf{z}_l | \mathbf{z}_{l+1}) = \mathcal{N}\left(z_l; \mathbf{f}_l^{\mathrm{gen}}(\mathbf{z}_{l+1}), \mathbf{D}_l^{\mathrm{gen}}\right) \tag{58}$$

where $\mathbf{f}_l$, can be arbitrarily complex, and in fact can contain (deterministic) neural networks in its own right. We describe the exact form for these functions below. Finally, the final layer generates binarised images,

$$\mathrm{P}(\mathbf{x}|\mathbf{z}_1) = \mathrm{Bernoulli}\left(\mathbf{x}; \sigma\left(\mathbf{f}_0^{\mathrm{gen}}(\mathbf{z}_1)\right)\right), \tag{59}$$

where the sigmoid function $\sigma(x) = 1/(1 + e^{-x})$ is applied element-wise. The recognition model is similar,

$$\mathrm{Q}(\mathbf{z}_l|\mathbf{z}_{l-1}) = \mathcal{N}\left(\mathbf{z}_l; \mathbf{f}_l^{\mathrm{rec}}(\mathbf{z}_{l-1}), \mathbf{D}_l^{\mathrm{rec}}\right) \tag{60}$$

with $\mathbf{z}_0 = \mathbf{x}$.

Given these definitions, it is trivial to differentiate $\mathcal{L}$ w.r.t. the parameters (i.e. the functions $\mathbf{f}_i$ and $\mathbf{g}_i$, and the noise levels, $\mathbf{D}_i^{\mathrm{rec}}$ and $\mathbf{D}_i^{\mathrm{gen}}$), thus, all we have yet to specify is the exact forms for $\mathbf{f}_i$ and $\mathbf{g}_i$. However, here we are using multiple layers of stochastic variables, whereas the usual set-up (under which we see the remarkable effectiveness of VAE's) is to have one stochastic layer, with many intermediate deterministic layers embodied in the functions $\mathbf{f}^{\text{gen}}$ and $\mathbf{f}^{\text{rec}}$ [24, 25]. Moreover, even work specifically investigating the possibility of using multiple stochastic layers still requires intermediate deterministic layers in $\mathbf{f}$ to achieve reasonable performance (e.g. [67]). Coincidentally, single biological neurons possess dendritic nonlinearities that make them more adequately modelled as two-layer neural networks [68]. Thus, we consider the activation functions (in both our generative and recognition networks) as arising from the combination of a number (40) of dendritic activations,

$$f_i(\mathbf{z}) = b_i + \sum_d W_{id} \left[ b_{id} + \sum_j W_{idj} \left[ z_j \right]_+ \right]_+ \tag{61}$$

where $i$ indexes the output cell, $d$ indexes the dendrite, $j$ indexes the presynaptic cell, and we have neglected layer indices and recognition/generative model labels for brevity. The input from the previous layer enters through $\mathbf{z}$, and as this variable is Gaussian distributed, it could be positive or negative, and so is best interpreted as a membrane potential. To obtain a firing rate, we thus apply a nonlinearity to $z_i$, in particular, we use a rectified linear activation function, $[z_i]_+$. Furthermore, the term $b_{id} + \sum_j W_{idj} [z_j]_+$ represents the inputs to the dendrites, and a rectified linear function is applied representing the dendritic nonlinearity, before they are integrated at the cell soma.

## 3.7 Deriving a measure of sampling speed based on a sample-based estimate of the mean

We measure sampling speed using the mean-square error of a sample-based estimate of the mean, integrating over a long time-period. In particular, we have $\mathbf{z}(t)$ generated by a dynamical system, which is a slightly generalised version of Eq. (32),

$$\dot{\mathbf{z}} = -\frac{1}{\tau} \mathbf{\Pi} \mathbf{z} + \frac{1}{\sqrt{\tau}} \sqrt{\mathbf{Q}} \, \boldsymbol{\eta} \tag{62}$$

and where $\mathbf{z}(t)$ has fixed marginal,

$$\mathrm{P}(\mathbf{z}(t)) = \mathcal{N}\left(\mathbf{z}(t); \mathbf{0}, \mathbf{\Sigma}\right). \tag{63}$$

The empirical mean of $\mathbf{z}(t)$ is

$$\boldsymbol{\mu} = \frac{1}{T} \int_0^T dt \, \mathbf{z}(t). \tag{64}$$

Thus, the covariance of the empirical mean is,

$$\mathrm{Cov}\left[\boldsymbol{\mu}\right] = \frac{1}{T^2} \int_0^T dt \int_0^T dt' \mathrm{Cov}\left[\mathbf{x}(t), \mathbf{x}(t')\right] \tag{65}$$

where,

$$\mathrm{Cov}\left[\mathbf{x}(t), \mathbf{x}(t')\right] = \begin{cases} \mathbf{\Sigma} e^{-\frac{t'-t}{\tau} \mathbf{\Pi}^T} & \text{if } t < t' \\ e^{-\frac{t-t'}{\tau} \mathbf{\Pi}} \mathbf{\Sigma} & \text{if } t' < t. \end{cases} \tag{66}$$

Integrating over the region for which $t < t'$,

$$\frac{1}{T^2} \int_0^T dt \int_t^T dt' \text{Cov}\left[\mathbf{x}(t), \mathbf{x}(t')\right] \tag{67}$$

substituting for the covariance in this region,

$$= \frac{\mathbf{\Sigma}}{T^2} \int_0^T dt \int_t^T dt' e^{-\frac{t'-t}{\tau}\Pi^T}, \tag{68}$$

performing the outermost integral,

$$= \frac{\mathbf{\Sigma}}{T^2} \int_0^T dt \left[-\tau \Pi^{-T} e^{-\frac{t'-t}{\tau}\Pi^T}\right]_t^T \tag{69}$$

$$= \frac{\mathbf{\Sigma}\Pi^{-T}}{T^2} \int_0^T dt \left(\mathbf{I} - e^{-\frac{T-t}{\tau}\Pi^T}\right) \tag{70}$$

In the limit of large $T$, the first term in the integral dominates, and we obtain,

$$\approx \frac{\mathbf{\Sigma}\Pi^{-T}}{T}. \tag{71}$$

Performing the analogous integral for the other region ($t' < t$), we obtain,

$$\text{Cov}\left[\boldsymbol{\mu}\right] = \frac{1}{T}\left(\mathbf{\Sigma}\Pi^{-T} + \Pi^{-1}\mathbf{\Sigma}\right). \tag{72}$$

and as for any linear dynamical system, we can write $\Pi$ as,

$$\Pi = \left(\tfrac{1}{2}\mathbf{Q} + \mathbf{S}\right)\mathbf{\Sigma}^{-1}, \tag{73}$$

we can rewrite the covariance as,

$$\text{Cov}\left[\boldsymbol{\mu}\right] = \frac{1}{T}\mathbf{\Sigma}\left(\left(\tfrac{1}{2}\mathbf{Q} + \mathbf{S}\right)^{-T} + \left(\tfrac{1}{2}\mathbf{Q} + \mathbf{S}\right)^{-1}\right)\mathbf{\Sigma} \tag{74}$$

exploiting the symmetry of $\mathbf{Q}$ and the antisymmetry of $\mathbf{S}$, we obtain,

$$\text{Cov}\left[\boldsymbol{\mu}\right] = \frac{1}{T}\mathbf{\Sigma}\left(\left(\tfrac{1}{2}\mathbf{Q} - \mathbf{S}\right)^{-1} + \left(\tfrac{1}{2}\mathbf{Q} + \mathbf{S}\right)^{-1}\right)\mathbf{\Sigma} \tag{75}$$

multiplying and dividing, such that both terms have the same "denominator"

$$, \text{Cov}\left[\boldsymbol{\mu}\right] = \frac{1}{T}\mathbf{\Sigma}\left(\tfrac{1}{2}\mathbf{Q} - \mathbf{S}\right)^{-1}\left(\left(\tfrac{1}{2}\mathbf{Q} - \mathbf{S}\right) + \left(\tfrac{1}{2}\mathbf{Q} + \mathbf{S}\right)\right)\left(\tfrac{1}{2}\mathbf{Q} + \mathbf{S}\right)^{-1}\mathbf{\Sigma}, \tag{76}$$

simplifing the "numerator",

$$= \frac{1}{T}\mathbf{\Sigma}^{1/2}\left(\mathbf{Q} - \mathbf{S}\right)^{-1}\mathbf{Q}\left(\mathbf{Q} + \mathbf{S}\right)^{-1}\mathbf{\Sigma}^{1/2}. \tag{77}$$

and in our case, we assume the noise covariance is proportional to the identity, $\mathbf{Q} = \sigma^2 \mathbf{I}$ giving,

$$= \frac{\sigma^2}{T}\mathbf{\Sigma}\left(\sigma^4 \mathbf{I} - \mathbf{S}^2\right)^{-1}\mathbf{\Sigma}. \tag{78}$$

Finally, we note that as the intention is to measure sampling speed, we should ensure that we have normalized for the effect of the underlying covariance (i.e. if the covariance is very large, this will inevitably reduce the precision of a sample-based estimate of the mean). As such, we use (the trace of),

$$\mathbf{\Sigma}^{-1/2}\text{Cov}\left[\boldsymbol{\mu}\right]\mathbf{\Sigma}^{-1/2} = \frac{1}{T}\mathbf{\Sigma}^{1/2}\left(\mathbf{Q} - \mathbf{S}\right)^{-1}\mathbf{Q}\left(\mathbf{Q} + \mathbf{S}\right)^{-1}\mathbf{\Sigma}^{1/2}, \tag{79}$$

or, with isotropic noise,

$$\mathbf{\Sigma}^{-1/2}\text{Cov}\left[\boldsymbol{\mu}\right]\mathbf{\Sigma}^{-1/2} = \frac{\sigma^2}{T}\mathbf{\Sigma}^{1/2}\left(\sigma^4\mathbf{I} - \mathbf{S}^2\right)^{-1}\mathbf{\Sigma}^{1/2}. \tag{80}$$

Finally, we use the trace of the above quantity as a measure of the mean-squared error in sample-based estimates of the mean, and thus of sampling speed.

Now, it is interesting to consider the significance of this quantity in more depth. In particular, one way to form a dynamical-system sampler with good properties is to set $\mathbf{S} = \mathbf{0}$, and match the process noise to the stationary covariance, $\mathbf{Q} = \mathbf{\Sigma}$, in which case the N-dimensional dynamical system effectively reduces to $N$ 1-dimensional dynamical systems, with equal sampling speeds in each direction. If we substitute $\mathbf{S} = \mathbf{0}$ and $\mathbf{Q} = \sigma^2\mathbf{\Sigma}$ into Eq. (77) we obtain,

$$\mathbf{\Sigma}^{-1/2}\text{Cov}\left[\boldsymbol{\mu}\right]\mathbf{\Sigma}^{-1/2} = \frac{1}{T\sigma^2}\mathbf{I} \tag{81}$$

i.e. the more noise we inject, the faster we sample. Thus, to get some intuition for our case with non-zero $\mathbf{\Sigma}$, and isotropic noise, we can try setting the mean estimate to be uncorrelated, and to have the same variance in each direction,

$$\mathbf{\Sigma}^{-1/2}\text{Cov}\left[\boldsymbol{\mu}\right]\mathbf{\Sigma}^{-1/2} \propto \frac{1}{T}\mathbf{I} \propto \frac{\sigma^2}{T}\mathbf{\Sigma}^{1/2}\left(\sigma^4\mathbf{I} - \mathbf{S}^2\right)^{-1}\mathbf{\Sigma}^{1/2} \tag{82}$$

which requires,

$$\mathbf{\Sigma} \propto \sigma^4\mathbf{I} - \mathbf{S}^2 \tag{83}$$

and solving for $\mathbf{S}$ gives,

$$\mathbf{S}^2 = -q\left(\frac{1}{\alpha}\mathbf{\Sigma} - q\mathbf{I}\right) \tag{84}$$

intuitively, tells us that as the process noise decreases (i.e. $q \to 0$), the optimal antisymmetric matrix, $\mathbf{\Sigma}$, should simply be the square root of minus the stationary covariance, $\mathbf{\Sigma}$. More formally, note that for $\mathbf{S}$ to be antisymmetric, we need $\mathbf{\Sigma}/\alpha - q\mathbf{I}$ to be positive definite, i.e. we need $\alpha$ to be greater than some lower-limit. This makes sense, because for a fixed $q$, the slowest possible sampling rate is above zero. This is a really nice approach, as it takes us directly to the right answer. However, finding an objective function that actually takes us to this location is likely to be decidedly non-trivial.

### 3.8 Obtaining a biologically plausible regulariser on S

If we simply optimize a sampling-speed objective, there is a risk that $\mathbf{S}$ could grow unboundedly large. To avoid this possibility, we need a regularizer. While typical regularizers might be placed directly on the weights, $\Pi$, we consider this inappropriate, as it does not take into account the degree to which each weight is used in practice (i.e. the firing rate or variance of the presynaptic cell). To do this, we consider regularizing the dynamics matrix, multiplied by the stationary covariance. This has the advantage of having a "parameter free" interpretation, as,

$$\Pi \boldsymbol{\Sigma} = \Pi \mathrm{E}\left[\mathbf{x}\mathbf{x}^T\right] = \mathrm{E}\left[\dot{\mathbf{x}}\mathbf{x}^T\right], \tag{85}$$

and also as depending only on the process noise and skew-symmetric matrix, $\mathbf{S}$ (so that this regularizer does not risk corrupting the stationary distribution),

$$\Pi \boldsymbol{\Sigma} = \tfrac{1}{2}\mathbf{Q} + \mathbf{S}. \tag{86}$$

In particular, we use the Frobenius norm of this quantity, with a small (0.01) multiplicative constant.